\pdfoutput=1 % only if pdf/png/jpg images are used
\RequirePackage{ifpdf}
\documentclass{JINST}

\RequirePackage{xspace}
\newcommand{\CfourHten}{\ensuremath{\mathrm{C}_4\mathrm{H}_{10}}\xspace}
\newcommand{\He}{\ensuremath{\mathrm{He}}\xspace}
\newcommand{\HeIso}{\ensuremath{\He:\CfourHten}\xspace}

\title{\bf Study of the single cluster response of a helium-isobutane drift chamber prototype using 8~keV X-rays}

\author{G.~Cavoto$^a$,
S.~Dabagov$^b$,
D.~Hampai$^c$,
G.~Piredda$^a$,
F.~Renga$^a$\thanks{Corresponding author.}~,
E.~Ripiccini$^{a,d}$
and C.~Voena$^a$\\
\llap{$^a$}INFN Sezione di Roma, Piazzale A. Moro 2, 00185 Roma, Italy\\
\llap{$^b$}P.~N.~Lebedev Phys. Inst. and NRNU MEPhI, Moscow, Russia\\
\llap{$^c$}INFN Laboratori Nazionali di Frascati, Via E. Fermi 40, 00044 Frascati, Italy\\
\llap{$^d$}Dipartimento di Fisica dell'Universit\`a ``Sapienza'', Piazzale A. Moro 2, 00185 Roma, Italy\\
E-mail: \email{francesco.renga@roma1.infn.it}}

\abstract{The identification of single clusters in the electronic signals produced by ionizing particles within a drift chamber is expected to
significantly improve the performances of this kind of detectors in terms of particle identification capabilities and space resolution. 
In order to develop refined cluster recognition algorithms, it is essential to measure the response of the chamber and its electronics to
single ionization clusters. This can be done by irradiating the chamber with X-rays. We report here on the studies performed on a 
drift chamber prototype for the MEG-II experiment at the X-ray facility of the INFN Frascati's National Laboratories ``XLab Frascati''. 
The prototype is operated with a helium-isobutane mixture and instrumented with high bandwidth custom pre-amplifiers.
The results of this study have been used to develop an innovative method for cluster recognition, based on the Wiener filter technique. 
As a side measurement, we also performed a study of the gas gain in a configuration which is similar to that of the MEG-II experiment.}

\keywords{Wire Chambers; Ionization and excitation processes; Charge transport and multiplication in gas; cluster finding, calibration and fitting methods; Digital signal processing (DSP); Performance of High Energy Physics Detectors}

\begin{document}

%\tableofcontents

\section{Introduction}
\label{sec:intro}

In the development of gaseous detectors, and, in particular, of drift chambers, it is necessary to study the generation of the signals produced by ionizing particles in the adopted
gas admixture and electric field configuration, and to determine the response of the front-end electronics to these signals. This is crucial in order to validate the design of the
chamber, and to determine its expected performances through reliable simulations. An X-ray source, used to irradiate a drift chamber prototype reproducing the design of the
final detector, is the ideal tool to fulfill this task.

Charged particles ionizing a gas admixture produce clusters of a few electron-ion pairs at different locations along their path. In a drift chamber, the electrons in a cluster 
drift together toward an anode wire, where an ionization avalanche occurs, with a characteristic multiplication factor per electron (\emph{gas gain}). In this way, each cluster
produces an individual signal in the chamber (\emph{single-cluster signal}), but the observed waveform will be indeed a superposition of several single-cluster signals.

\begin{figure}[!h]
\centering
\includegraphics[width=70mm]{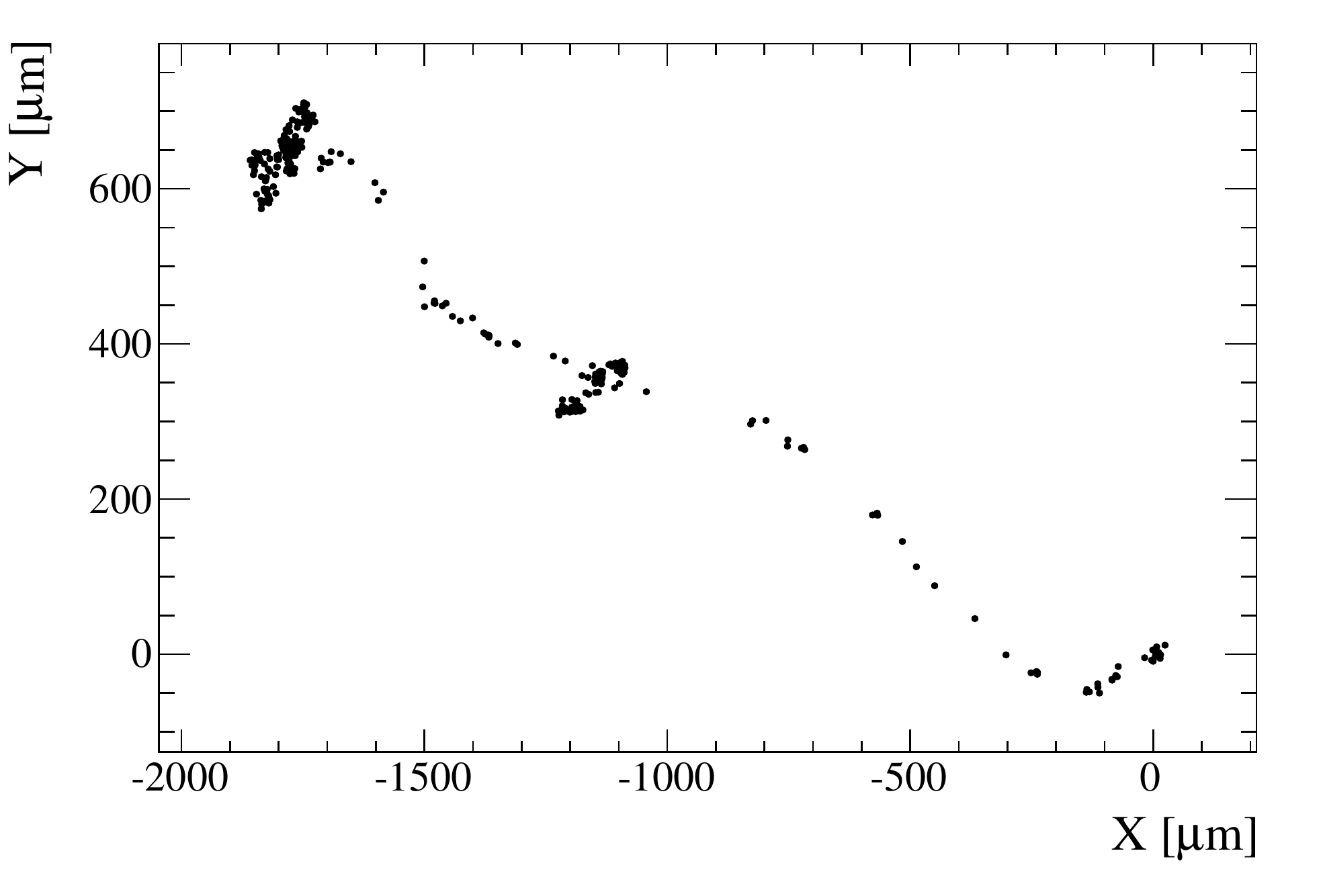}
\includegraphics[width=70mm]{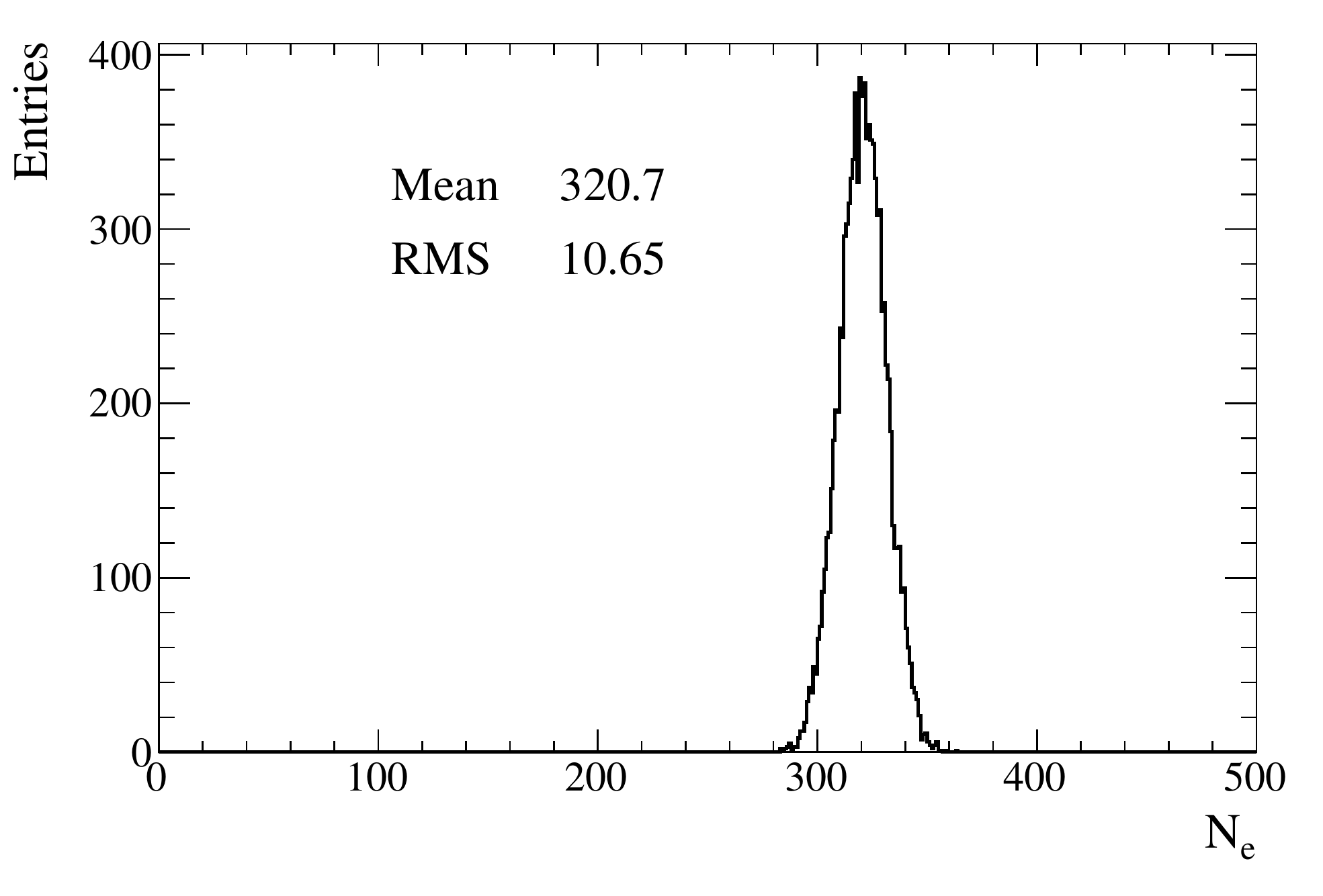}
\caption{Results of a DEGRAD simulation in a \HeIso (89:11) mixture. Left: position of the electrons produced in a typical event by a single 8.5 keV X-ray ionizing the gas at the 
origin of the reference system. Right: distribution of the number of electrons for a set of events.}
\label{fig:XrayIon}
\end{figure}

Conversely, photo-electric effect by an X-ray produces a single electron, which carries, in the first approximation, all the energy of the incident X-ray. This electron having a few 
keV kinetic energy flies at most a few hundred microns in a typical drift chamber gas admixture, before loosing all the energy by further ionizations, and eventually thermalizes. 
It results into a cluster of electron-ion pairs localized within $\sim1$~mm of the X-ray interaction point. The average number of pairs is also proportional to the X-ray energy. 
We used the DEGRAD software~\cite{degrad} to simulate these processes, assuming a \HeIso (89:11) mixture. Figure~\ref{fig:XrayIon} shows the position of the electrons produced 
in a typical event, and the distribution of the number of electrons produced by 8.5 keV X-rays.
Although the total extent of the electron cluster can be significantly large, most of the electrons tend to concentrate on small well-separated regions. Hence, in most cases, 
an X-ray will produce a signal very similar, in shape, to the single-cluster signals produced by a charged particle, although sometimes multiple clusters can show up. Moreover, 
the average integrated charge will be proportional to the average number of pairs (i.e. to the X-ray energy) and to the gain of the chamber. In conclusion, measuring both the charge and 
shape of the signals produced by X-rays allows measuring the gain of the chamber (if the X-ray energy spectrum is known) and determining the shape of single-cluster signals.
These two inputs allow, for instance, the front-end electronics to be optimized, the working conditions of the chamber to be determined (in order to avoid significant aging due to the large
amount of collected charge), and reliable simulations to be developed. This will be of paramount importance for the development of the Monte Carlo simulation for the MEG-II experiment.

The knowledge of the signal shape is even more important in the development of drift chambers with single-cluster detection capabilities, which are expected to significantly improve 
the performances of this kind of detectors by providing the number of ionization clusters for particle identification purposes (\emph{cluster counting})
and the drift time of each single cluster for position reconstruction purposes (\emph{cluster timing})~\cite{ClusCount,ClusTime}. Indeed, refined algorithms can be developed 
for the identification of clusters in the signal waveforms, if the single-cluster signal shape is well known.

We exploited the X-ray test facility (XLab Frascati)~\cite{Xlab} of the INFN Frascati's National Laboratories (LNF) to study the signals of a drift chamber prototype. The collected data 
allowed to develop new cluster identification algorithms, based on the Wiener filter technique~\cite{Wiener}, and study the cluster counting capabilities of our device. 

\section{The drift chamber prototype}

The prototype under test, built as an R\&D project for the MEG-II experiment~\cite{UpgradeProposal}, is described in detail in~\cite{reso_paper}. 
It is constituted by a gas-tight Aluminum body of 20~$\times$~20~$\times$~50~cm$^3$. 
The lateral faces are made of 1.5mm thick Al plates and/or 50~$\mu$m thick aluminized kapton windows, depending on the set-up. The other two faces are closed by 
two end-plates made of Au-plated Al and drilled to accomodate the feedthroughs where the wires were soldered and the gas pipe fittings.
The hole mask defines a 8 $\times$ 8 array of 7~mm-side square cells, each with a sense wire 
(25~$\mu$m Au-plated Tungsten) in the middle and 8 field wires (80~$\mu$m Au-plated Tungsten wires) all around. The prototype is instrumented with custom large-bandwitdh
preamplifiers in order to improve its single cluster recognition capabilities.

%---------------------------------------------------------------------------------------------------------------------
\section{Single cluster and gain measurement at XLab Frascati}

The XLab Frascati has a 4~m$^3$ shielded cabinet to host the detector or the sample to be irradiated. Inside the cabinet, an Oxford Apogee 40~W X-ray tube with Cu target is placed
that can be moved horizontally and vertically with millimetric precision. The drift chamber prototype has been positioned inside the cabinet so that the X-rays impinge 
perpendicularly on one of the long sides and enter the chamber through the 50~$\mu$m aluminized Kapton window, which has been mounted to avoid an excessive X-ray attenuation.
The vertical position of the tube with respect to the cell has been scanned to maximize the event rate.
The high voltage was supplied to the sense wire of the irradiated cell and its surrounding wires. %questo e' vero nelle misure che abbiamo usato.MA VERAMENTE?, NON CAPISCO
The data have been taken with both a 6-channel and a 1-channel versions of the pre-amplifier.
The amplified signal from the irradiated cell is readout with a high bandwidth differential probe to a 1~GHz bandwidth, 10~GS/s oscilloscope (LeCroy WaveRunner 610Zi).
% da qualche parte dobbiamo dire quanti dati abbiamo preso in ogni configurazioe.
%questo viene detto dopo.
% Dobbiamo aggiungere lo studio di run di fondo?
%non lo so, ma forse si

\subsection{The X-ray source}

\begin{figure}[!h]
\centering
\includegraphics[width=90mm]{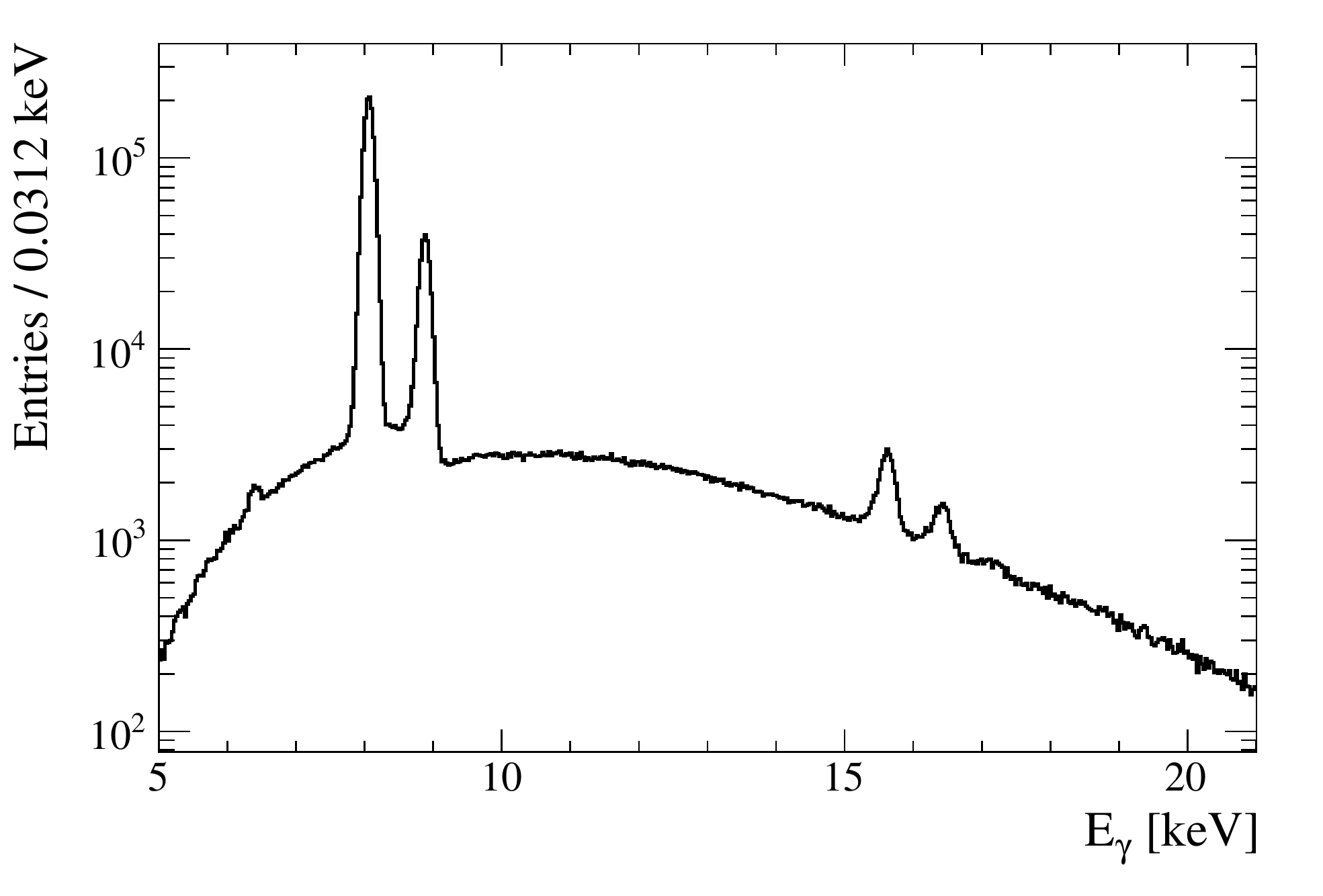}
\caption{Emission spectrum of the X-ray tube used at XLab Frascati, under the operating conditions adopted during the drift chamber prototype tests. The first peak results from the
superposition of the two $K_{\alpha}$ lines.}
\label{fig:dariush}
\end{figure}

The voltage of the source has been chosen to be 25~kV which provides good Cu emission lines ($K_{\alpha1}$ = 8.04778~keV, $K_{\alpha2}$ = 8.02783~keV, $K_{\beta1}$ = 8.90529~keV) over 
a broad background, as shown in Figure~\ref{fig:dariush}. The current of the tube can be varied to regulate the X-ray intensity. It was scanned between 5 and 25~$\mu$A, but most 
of the measurements were done at 5~$\mu$A, which is assumed through the paper unless explicitly stated. The intensity of the source at the origin was $\sim10^8$~$\gamma$/s. 
Taking into account the attenuation in air, kapton and gas before reaching the instrumented cell, and the expected ionization probability in the 7 mm of gas within the cell, a signal 
rate of about 10-100~kHz was expected.

\subsection{Single cluster signals}
\label{sec:siclu}

In the coaxial approximation, the current signal produced in a drift chamber by the avalanche initiated by a single electron has the characteristic form~\cite{Rolandi}
\begin{equation}
\label{eq:theo_current}
I(t) \propto \frac{1}{t+t_0} \, ,
\end{equation}
where the time constant $t_0$ depends on the ion mobility $\mu$ and the electrical configuration of the drift cell. In particular,
\begin{equation}
\frac{1}{t_0} = \frac{\mu}{a^2\pi\epsilon _0}\left(\sum_{m=1}^N c_{nm}U_m \right) \, ,
\end{equation}
where $a$ is the sense wire radius, $N$ is the number of wires, $c_{nm}$ is the capacitance matrix of the wire grid and $U_{n}$ the voltage of the wires. 
When the ions produced in the avalanche reach the field wires, after a time $t_{max}$, the signal stops. Hence, the condition
\begin{equation}
\label{eq:qtot}
Q_{tot} \equiv \int_0^{\infty} I(t) \, dt = \int_0^{t_{max}} I(t) \, dt = G \cdot q_e 
\end{equation}
will be satisfied, and it defines the normalization of $I(t)$ in Eq.~(\ref{eq:theo_current}), being $G$ the gas gain and $q_e$ the electron charge. The mobility for several 
ions in helium can be found in literature~\cite{Mobility}; although a 
reference for the mobility of isobutane ions in helium could not be found, according to the available measurements a reasonable range for our setup is 
10-20 $\rm{cm}^2\rm{V}^{-1}s^{-1}$. The capacitance matrix has been estimated by means of analytical calculations within GARFIELD.
As a result, we obtain $t_0$ = 0.150-0.300~ns and $t_{max}$ = 8-16~$\mu$s for a voltage of 1560 V.
Considering that the chamber has been operated within a $\pm$~10\% voltage range around this value, and $\mu$ is expected to be relatively 
stable with respect to the isobutane fraction, we do not expect significant differences of this shape through the configurations under test. 

The current signal is then amplified by the front-end electronics, resulting in a voltage signal
\begin{equation}
\label{eq:ItoV}
V(t) = \int_0^{t_{max}} w(t - t') \, I(t') \, dt \, ,
\end{equation}
where $w(t - t')$ is the electronics response to an infinitely fast unit charge signal (\emph{delta response}).

To measure the response of the chamber (including the electronics contribution) to the X-rays single clusters an average of the acquired signals has been performed, by using
a range of 100~ns (1000 sampling bins) around the signal, after having rescaled them to unit amplitude and aligned their starting time, taken at the half-maximum of the 
leading edge. For this study, we removed signals that saturated the oscilloscope acquisition ($\sim680$~mV) and signals whose full-width at half-maximum 
indicated the presence of multiple clusters. Moreover, to further remove a small fraction of noisy signals, waveforms deviating from the average signal by more than 
4 standard deviations in more than 10 sampling bins were also removed. The results were proved to be stable with respect to small changes of the selection criteria.

\begin{figure}[!h]
\centering
\includegraphics[width=70mm]{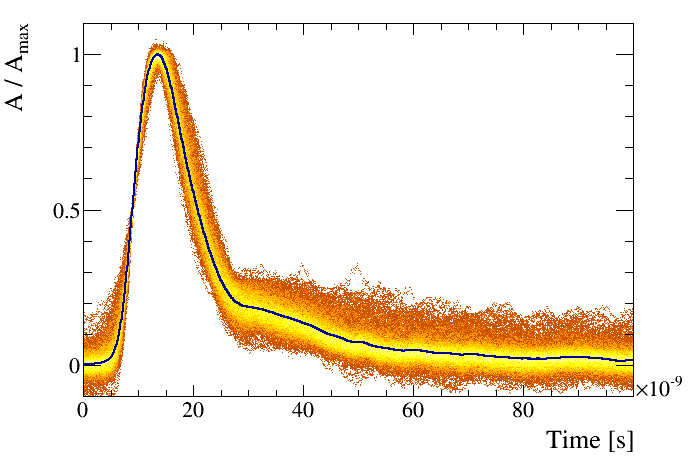}
\includegraphics[width=70mm]{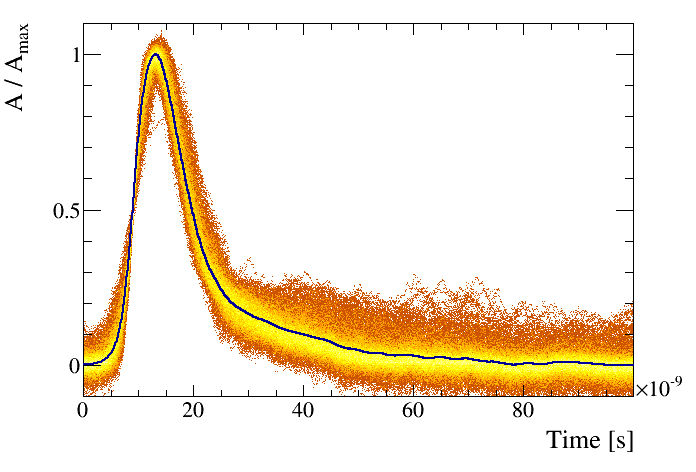}
\caption{Superposition of single cluster waveforms. Left: \HeIso (89:11) mixture with 1-channel pre-amplifier. Right: \HeIso (93:7) mixture with 6-channel amplifier.
Point density is larger in the lighter-colored regions. The dark line represents the average.}
\label{fig:shape}
\end{figure}

Figure~\ref{fig:shape} shows the superposition of the waveforms, normalized to their maximum amplitude. The dark line represents the average response. Two different configurations
have been tested:\HeIso (89:11) mixture with 1-channel pre-amplifier and \HeIso (93:7) mixture with 6-channel amplifier are shown.
%
%Figure~\ref{fig:shape3} shows the average response for different high-voltage/mixture/readout configurations.
%
%\begin{figure}[!h]
%\centering
%\includegraphics[width=90mm]{template_all.pdf}
%\caption{Average signal amplitude normalized to the maximum amplitude, of the single cluster response for different high-voltage/mixture/readout configurations.}
%\label{fig:shape3}
%\end{figure}
%
The shape is very similar, with small dependence on the high voltage and the mixture. Some difference is observed between the two pre-amplifiers, which 
could be due to a known impedance mismatch in the 1-channel preamplifier. The long tail expected from the slow ion drift is observed. The above response functions 
can be used in the MEG-II simulations and as signal templates in cluster counting/timing studies.
%----------------------------------------------------------------------------------------------------------

\subsection{Gain measurement} 

As a side measurement, we also tested a method for the gain measurement using X-rays. A reliable estimate of the gain is indeed crucial for the choice of the
working point of drift chambers operating in a high rate environment, as in MEG-II, in order to avoid intolerable aging effects due to an excessive charge collection.
 
Considering that the cluster produced by an X-ray contains an average of $N_e$ electrons, we can write the Gain $G$ as
\begin{equation} 
G = \frac{Q_{tot}}{N_e q_e} \, .
\end{equation}

The charge of the collected signals has to be estimated from the integral of the voltage signals in a 100~ns window. In this case, no selection is applied on the full-width half-maximum
in order to collect all the charge released in the event, while an algorithm based on the single cluster templates has been used to get a correct estimate 
of the charge also for the saturated signals.

\begin{figure}[!h]
\centering
\includegraphics[width=90mm]{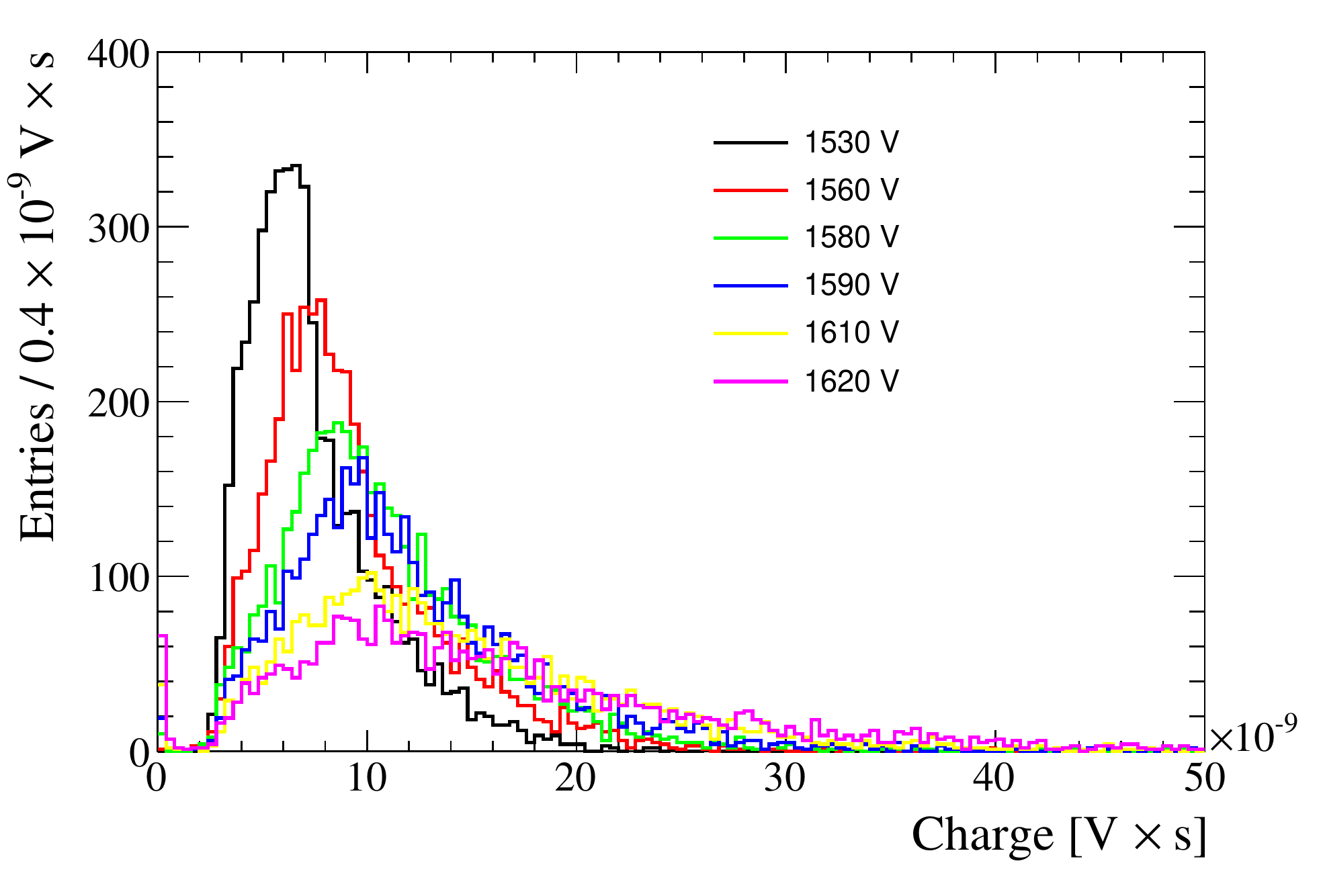}
\caption{Signal integrals $\mathcal{I}_{meas}$ in a 100~ns window for different high voltages. The mixture is \HeIso (89:11). The signals are readout with the 1-channel preamplifier.}
\label{fig:charge}
\end{figure}

Figure~\ref{fig:charge} shows the signal integrals for the % aggiungere spettro fondo
\HeIso (89:11) mixture %al momento e' inconsistente la label
with different high voltages and 1-channel preamplifier.
As expected, at higher high voltages, the average charge is higher.
The width of the distribution is also increasing with the high voltage as it increases with the gain.
%che dice il RB? In babar si vede che per G circa 4x10^5 la larghezza e' 30-40%
% il nostro problema e' che noi non ce l'abbiamo simmetrica babar si, sara'
% il fondo di raggi X di background?

\begin{figure}[!h]
\centering

\includegraphics[width=70mm]{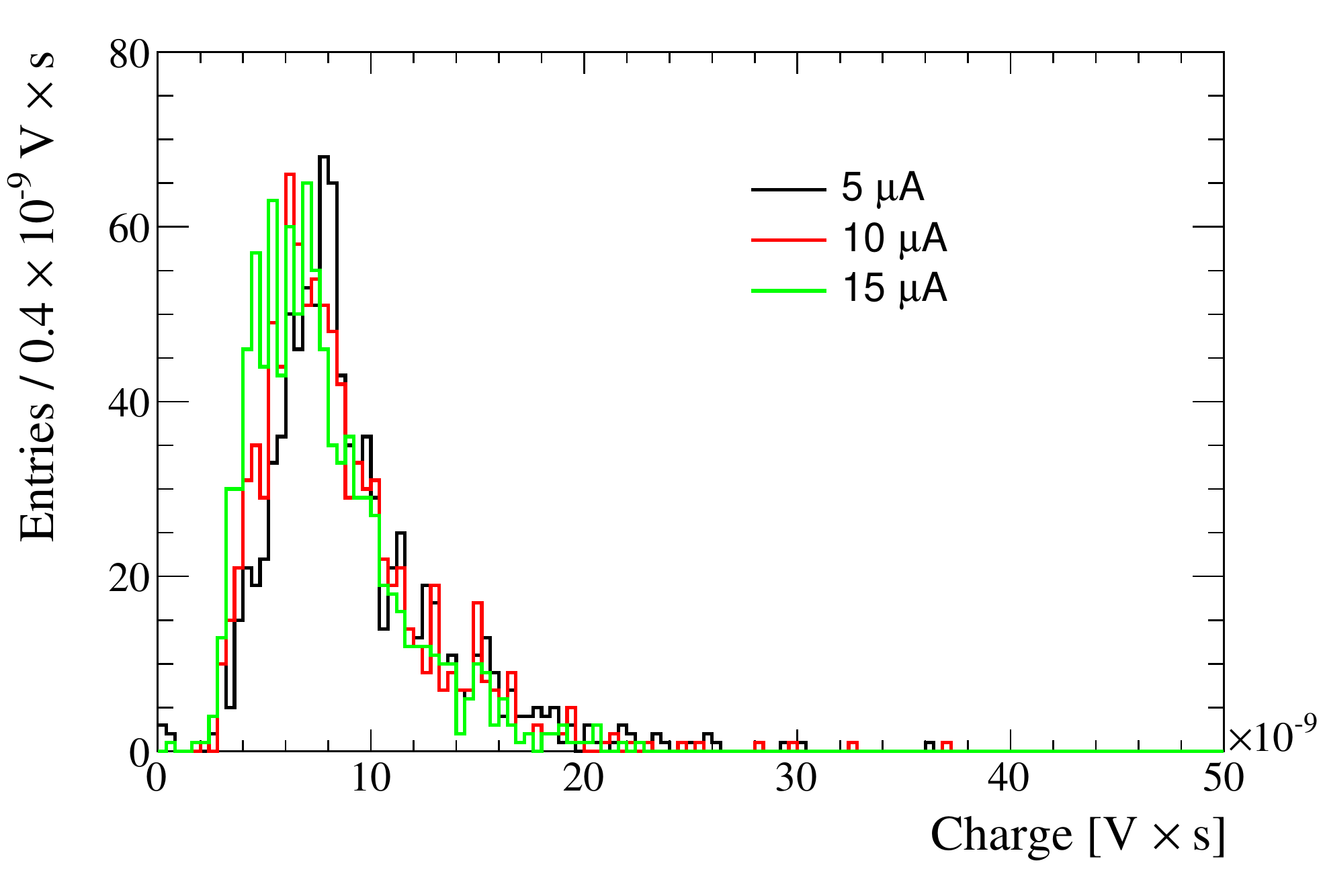}
\includegraphics[width=70mm]{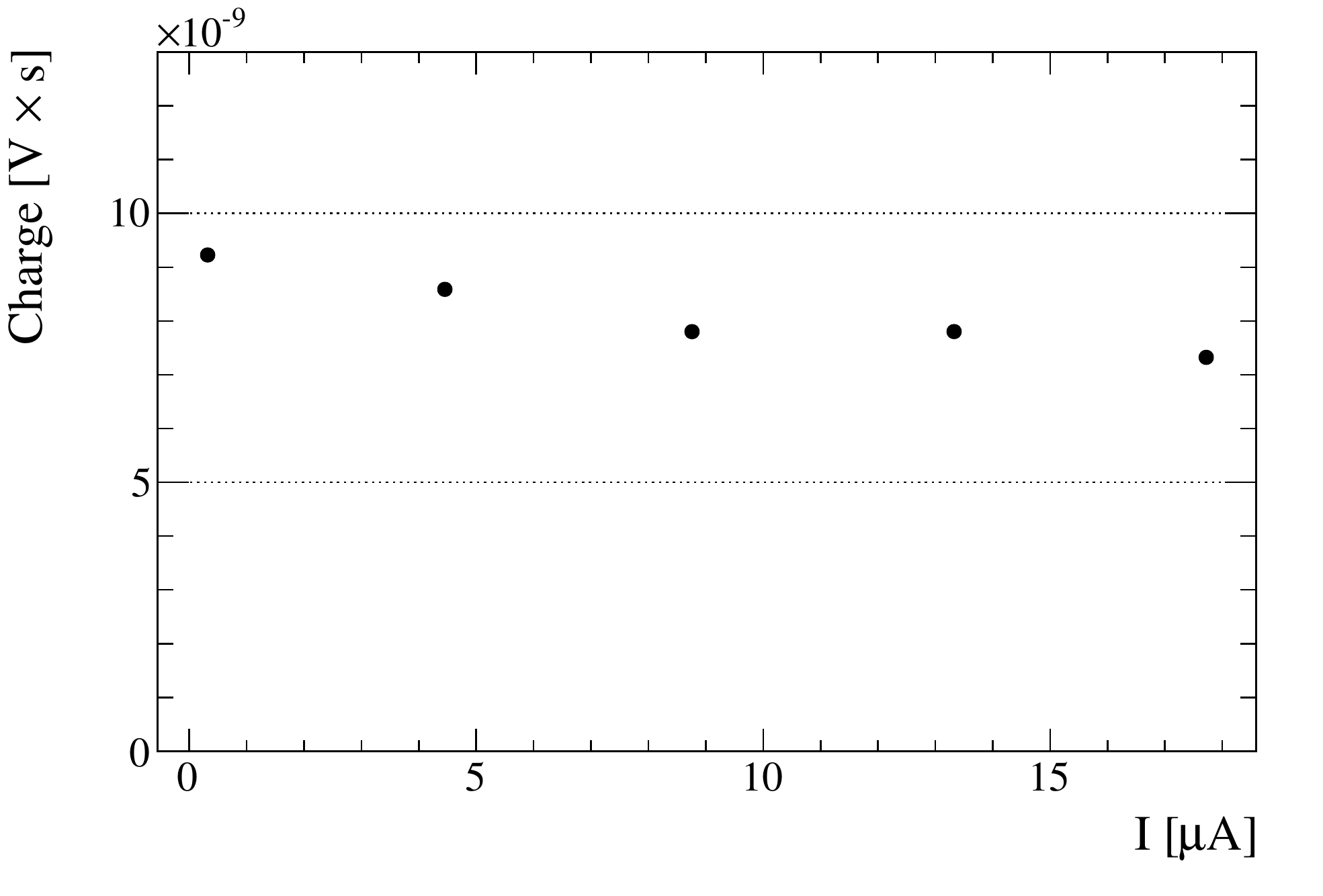}
\caption{Left: signal integrals $\mathcal{I}_{meas}$ (in V~$\times$~s) in a 100~ns window for the \HeIso (89:11) mixture, HV=1560~V, 1-channel preamplifier and 
different currents of the X-ray tube. Right: mean value of the integral distribution versus tube current.}
\label{fig:charge2}
\end{figure}

A scan of the tube current has been performed up to 25~$\mu$A, while the working point used for all other measurements is 5~$\mu$A. Figure~\ref{fig:charge2} shows the integrated 
charge distribution and the variation of the charge mean value as a function of the tube current. The mean value is corrected according to the expected gain vs. high voltage 
curves (see below), to take into account the voltage drop across the high voltage supply circuit of the prototype. Nonetheless, a decrease at higher tube current is observed, 
that may be due to space charge effects appearing at high rate. An exponential fit is used to extrapolate the integral to $I = 0$, and indicates a 9$\%$ gain 
suppression at 5~$\mu$A with respect to the zero-current limit.
%cosa e' la linea? Un fit? Abbiamo un aspettazione con cui confrontarci?
%

In order to extract a charge measurement from the signal integrals, two factors need to be considered: the response of the front-end electronics and the loss of the charge in the
long signal tails due to the limited integration window. Calculations based on the coaxial approximation of Sec.~\ref{sec:siclu} give an expected (43~$\pm$~4)$\%$ charge loss 
for $t_0$ ranging from 0.150 to 0.300~ns. 
From Eq.~(\ref{eq:ItoV}), the measured signal integral can be written as
\begin{eqnarray}
\mathcal{I}_{meas} &=& \int_0^T V(t) \, dt = \int_0^T dt \int_0^{t_{max}} w(t - t') I(t') \, dt' = \int_0^{t_{max}} dt' I(t') \int_0^{T-t'} dt'' w(t'') \, ,
\end{eqnarray}
with $T$ = 100~ns. Once the electronics response $w(t)$ is known, the integral can be performed numerically and the ratio of $Q_{tot}$ to $\mathcal{I}_{meas}$, which is
independent of $G$, can be used as a conversion factor to transform the signal integral into a collected charge. It accounts for both the limited integration window 
and the electronics response, which has been extracted
from a circuit simulation including the preamplifier, the drift chamber equivalent circuit and the
HV board. A conversion factor of $2.1 \times 10^{-3}$ ($3.0 \times 10^{-3}$) $\frac{C}{V \times s}$ has been found for the 1-channel (6-channel) pre-amplifier.

We also need to take into account the combined effect of the time distribution of the secondary electron clusters produced by the X-rays (Sec.~\ref{sec:intro}), the space 
charge felt by clusters reaching the sense wire with a significant delay and the oscilloscope trigger threshold  (124 mV). A waveform 
simulation, driven by the results of DEGRAD, GARFIELD, and a study of the space charge effect on waveforms produced by charged tracks, has been used 
to model these effects and revealed a 35 to 45$\%$ underestimate of the true gain, increasing with the gain itself.

\begin{figure}[!h]
\centering
\includegraphics[width=90mm]{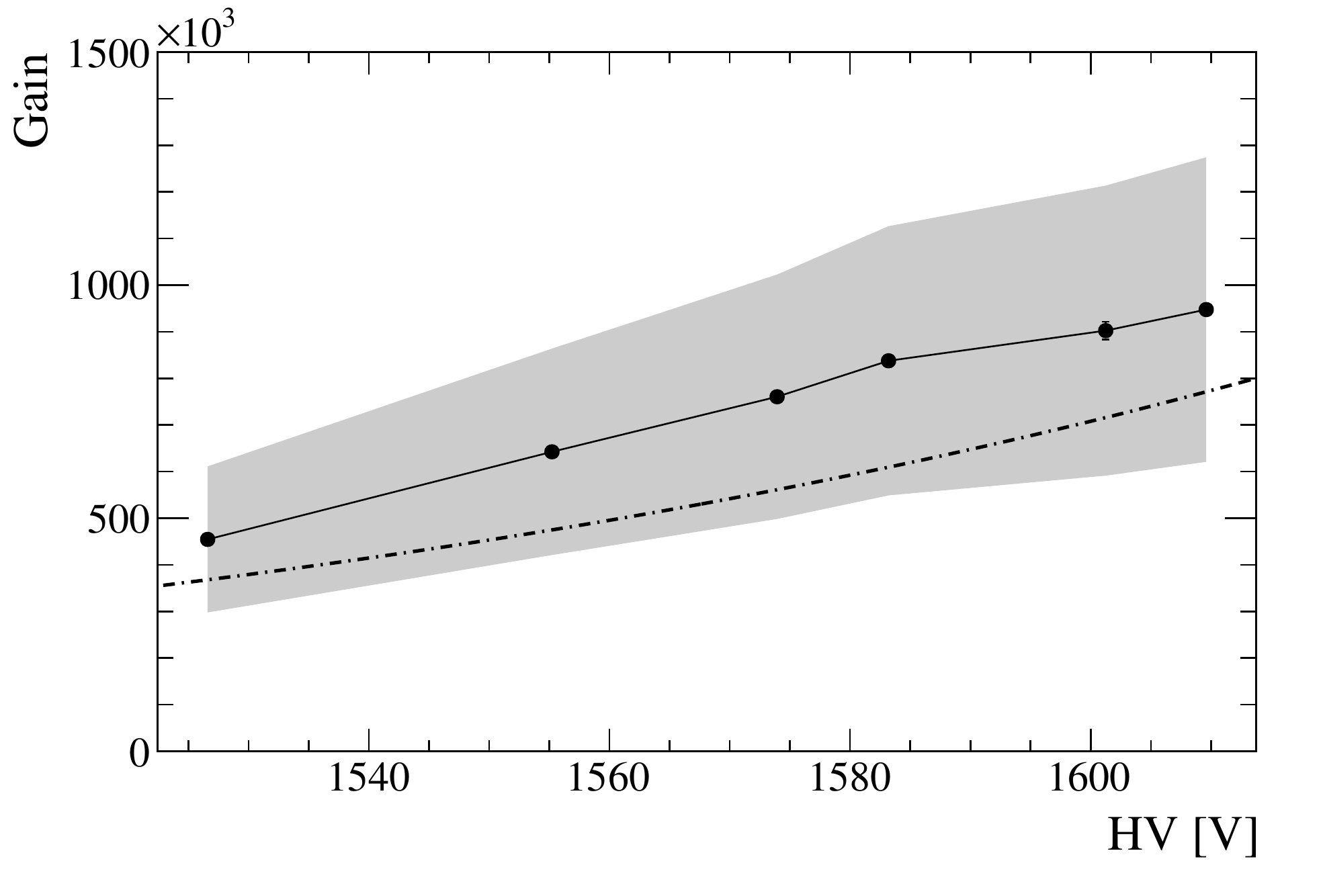}
\caption{Gas gain as a function of the drift chamber high voltage for the \HeIso (89:11) mixture. The gray belt shows the systematic uncertainties described in the text. 
The predictions of a GARFIELD simulation are given by the dashed line.}
\label{fig:gasgain}
\end{figure}

With all these calibration, a measurement of the average gain can be extracted from the signal integral distribution. For this purpose, the X-ray spectrum of Figure~\ref{fig:dariush} 
has been convolved with a Gaussian response function and multiplied by a threshold function, and then fitted to the measured spectra. Only the gain and the width of the
Gaussian response are floating in the fit. The fitted gas gain as a function of the high voltage is shown in Figure~\ref{fig:gasgain}. A 34$\%$ systematic error, fully correlated
among the data points, is included in the plot (gray belt). It accounts for the three dominant sources of uncertainty: the pre-amplifier calibration ($\pm$~30$\%$, estimated from 
the differences in the calibrated response of different amplifiers under the same data taking conditions), the correction factor from waveform 
simulations ($\pm$~15$\%$, estimated by variating the simulation inputs) and the $t_0$ value used to extract the expected charge loss ($\pm$~9$\%$). The results are also compared with a 
GARFIELD simulation which reproduces the geometrical and electrical configuration of the prototype, and takes into account the gas properties by performing a microscopic 
simulation of the avalanche process initiated by thermalized electrons. A satisfactory agreement is found.
\\

\section{A Wiener filter approach for cluster counting}

The Wiener filter technique is widely used for the deconvolution of a known delta response from a signal affected by noise. If the underlying signal is
assumed to be a Dirac delta-function, the result of the deconvolution is a narrow peak at the signal leading edge, and it makes possible to resolve near signals. Hence, 
this technique is promising also for searching single clusters in the signal produced by a charged particle within a drift chamber.

Given the noise spectrum and the expected single-cluster signal template, we consider a Wiener filter defined by the following response 
function in the frequency domain
\begin{equation}
H(\nu) = \frac{S^*(\nu)}{|S(\nu)|^2 + |N(\nu)|^2} \, ,
\end{equation}
where $S(\nu)$ and $S(\nu)$ are the Fourier transform of the single-cluster signal template and the average Fourier transform of the noise, 
respectively. Compared to the corresponding pure Fourier deconvolution, $H(\nu) = S^*(\nu)/|S(\nu)|^2$, the Wiener deconvolution efficiently suppresses the noise, 
but the peaks in the filtered waveform get wider. Conversely, compared to the optimum filter $H(\nu) = S^*(\nu)/|N(\nu)|^2$, the noise suppression in less efficient.
In order to optimize the filter for cluster identification, we adopted this response function as a compromise
\begin{equation}
H(\nu) = \frac{S^*(\nu)}{|S(\nu)|^2 + k^2|N(\nu)|^2} \, ,
\end{equation}
where $k$ is a factor to be optimized, while $S(\nu)$ is extracted considering the average signal amplitude. After applying the filter, a cluster emerges as a narrow 
peak at the leading time of the single-cluster signal, and can be identified looking for local maxima in the filtered waveform, separated from local minima by more than 
$n$ times the noise RMS measured in the filtered waveform itself, where $n$ has also to be optimized (a range of $[-50,+200]$~ns around the
leading edge of the signal is used to search for clusters).

\begin{figure}[!h]
\centering
\includegraphics[width=90mm]{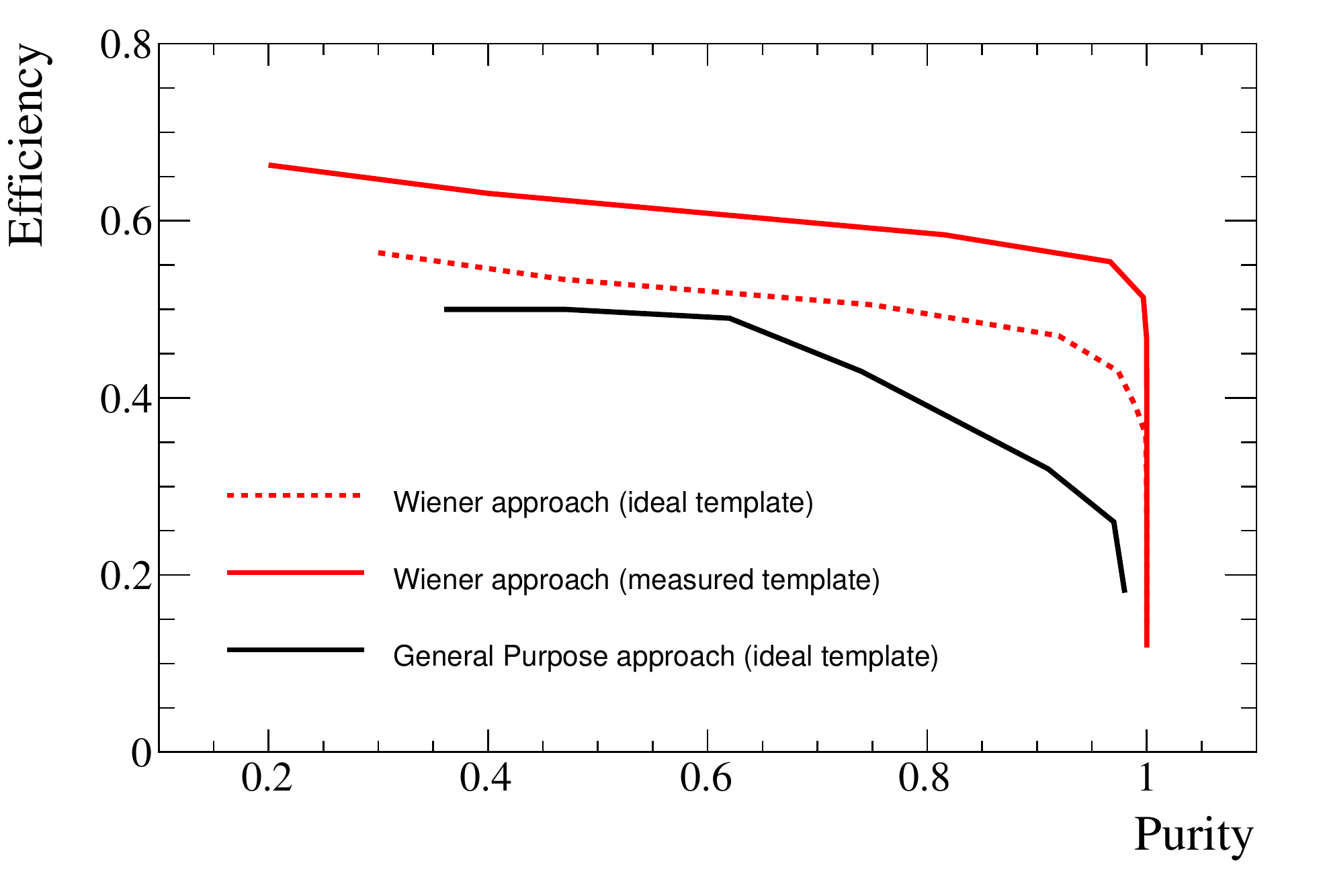}
\caption{Performances of the Wiener filter approach for simulated ideal signals (rise time $\sim$ 2~ns), compared to the best performances 
obtained from the general purpose peak-finding approach described in~\cite{TSpectrum}. The results of a simulation based on the measured
template and noise are also shown.}
\label{fig:perf_wiener}
\end{figure}

The Wiener filter method has been first applied to simulated waveforms in order to test its theoretical capabilities. In a first simulation we assumed a Gaussian, 2 mV RMS 
noise and an ideal signal shape with a very short rise time ($\sim$ 2~ns), a $1/t$ tail, and an average amplitude of 30~mV per cluster. Cluster times are
simulated according to the expectations for the geometry of our prototype. The performances of the algorithm are evaluated in terms of cluster recognition 
efficiency and probability of a cluster of not being a fake (\emph{purity}), and compared to the results obtained
with a general purpose peak-finding algorithm~\cite{TSpectrum}. Several combinations of $n$ and $k$ have been tried, and the best ones have been chosen for the 
comparison. The same approach is adopted to find the best combinations of parameters for the general purpose algorithm. The results are shown in Figure~\ref{fig:perf_wiener}. 
The Wiener filter approach, being tailored on the expected single cluster waveforms, clearly outperforms the general purpose algorithm.

In order to test the capabilities of the Wiener filter technique on real data, we exploited a set of data collected with our prototype at the
LNF Beam Test Facility (BTF)~\cite{reso_paper}.

\begin{figure}[!h]
\centering
\includegraphics[width=90mm]{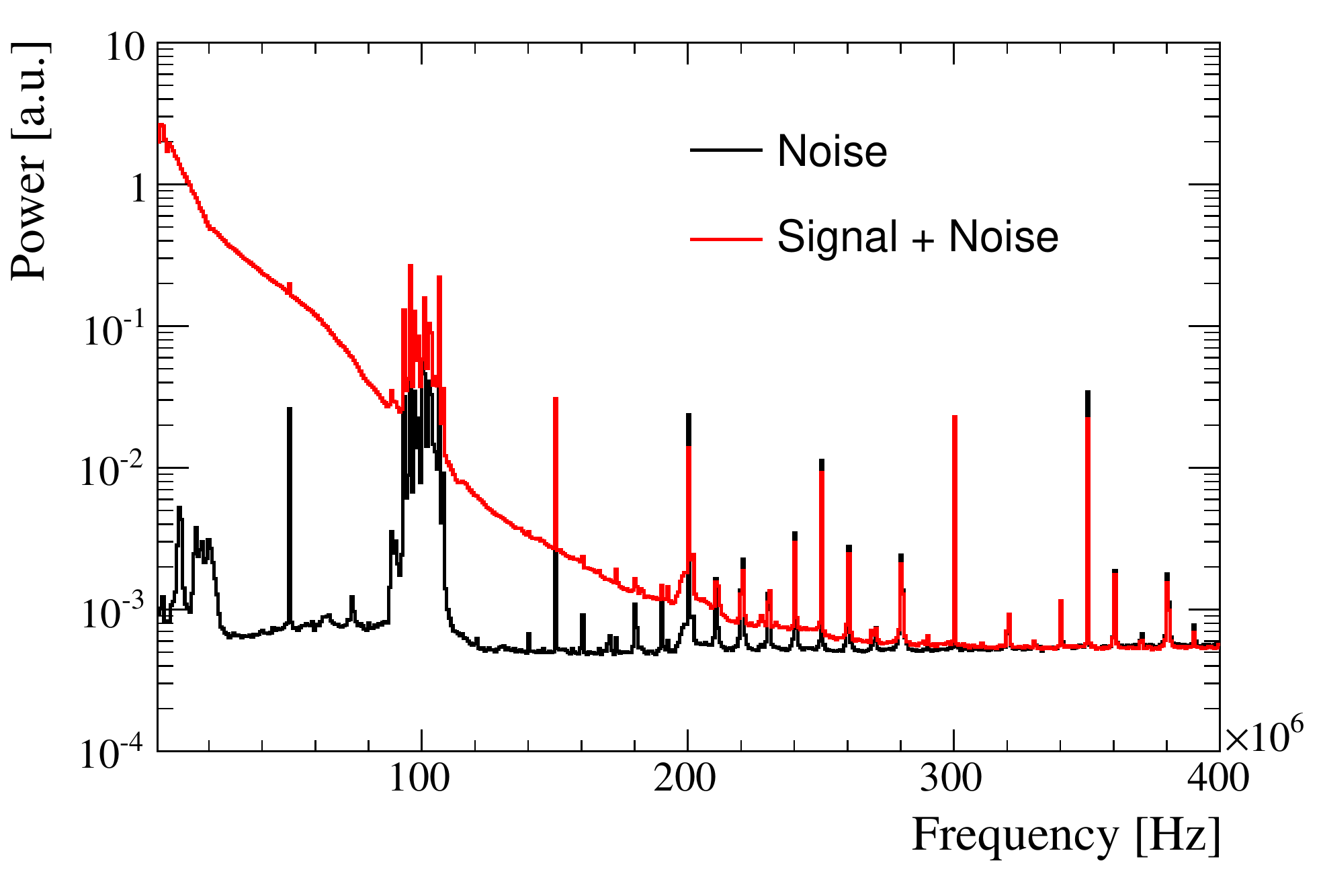}
\caption{Average frequency spectrum of the noise (black) and signal plus noise (red) waveforms at BTF.}
\label{fig:FT}
\end{figure}

\begin{figure}[!h]
\centering
\includegraphics[width=120mm]{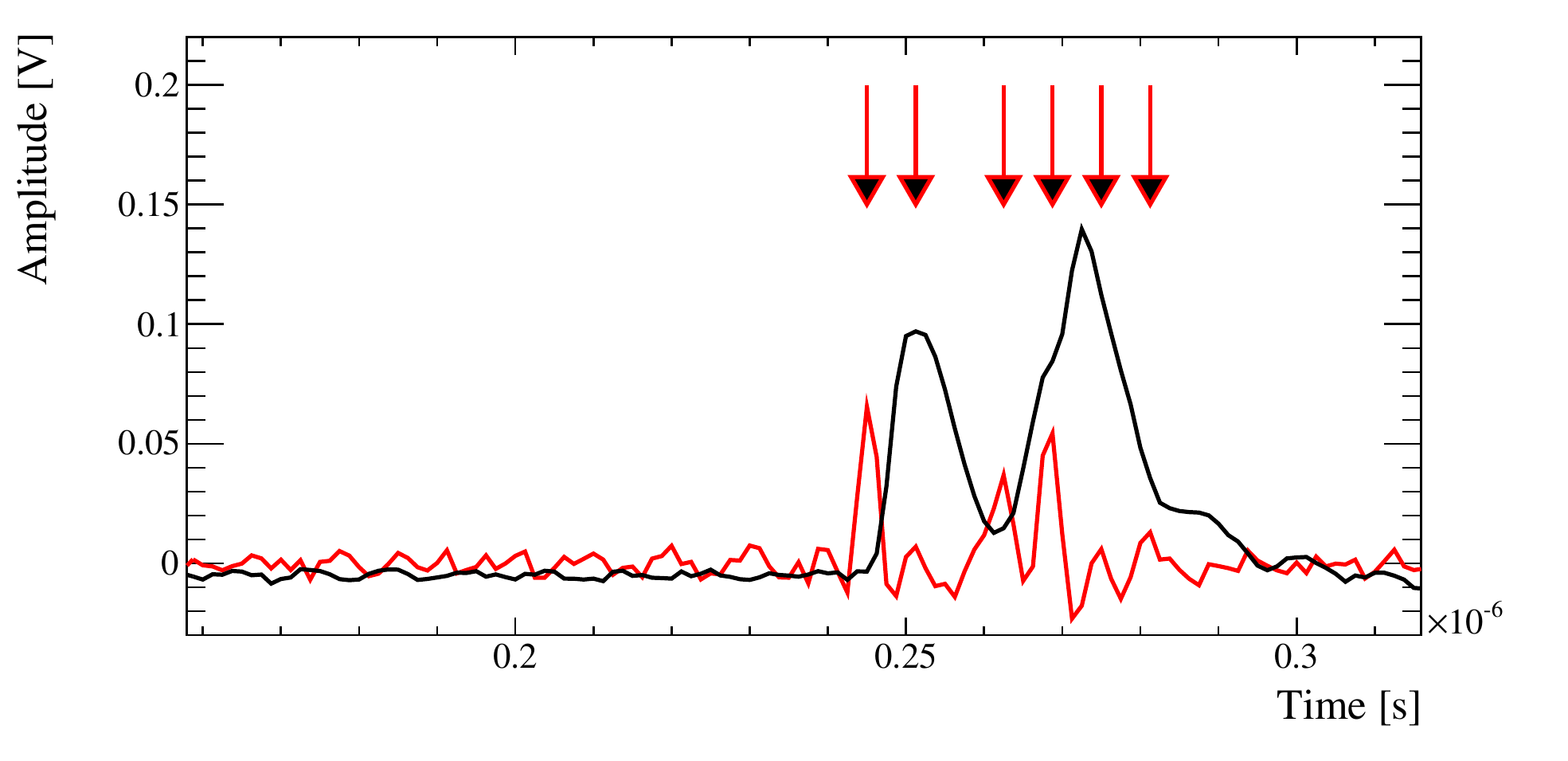}
\caption{Segment of a BTF signal waveform with the result of the Wiener filter super-imposed (red, arbitrarily normalized) and the identified clusters indicated by arrows. Notice that
the filter is built in such a way that peaks show up at the leading edge of the cluster signal.}
\label{fig:signal}
\end{figure}

Figure~\ref{fig:FT} shows the average noise spectrum and the average frequency spectrum of waveforms with a reconstructed hit. 
Thanks to the measurements performed at XLab Frascati, we can use the single cluster template to build the Wiener Filter, along with this noise
spectrum. The results of a simulation generated according to the signal template measured at XLab Frascati and the noise observed at BTF are also 
shown in Fig.~\ref{fig:perf_wiener}. The best performances are obtained for $k \sim 5$. Hence we used $k = 5$ to build the filter and $n = 4$ for cluster 
identification in the filtered waveform. Figure~\ref{fig:signal} shows an example of waveform with the identified clusters.

\begin{figure}[!h]
\centering
\includegraphics[width=90mm,scale=0.5]{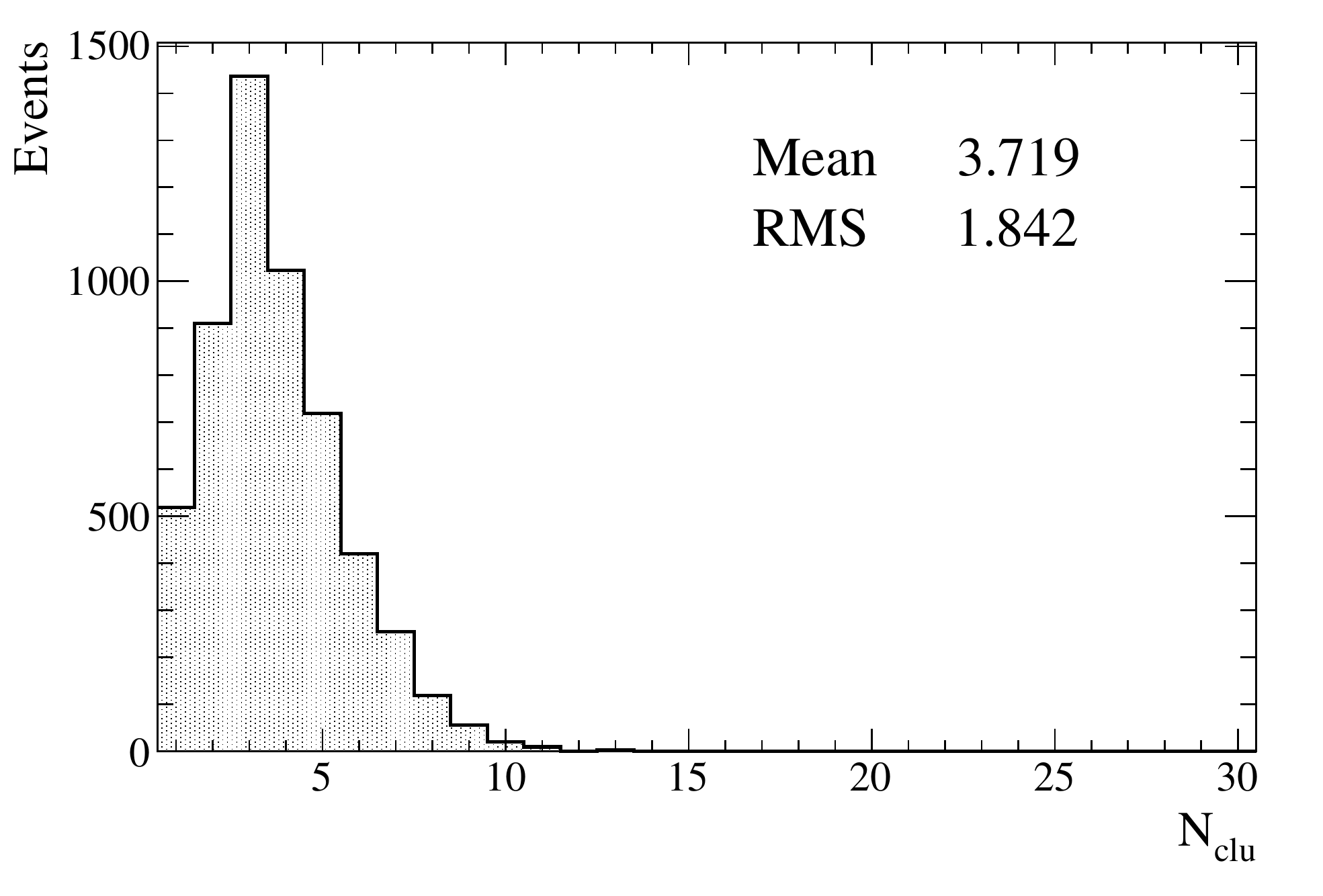}
\caption{Number of identified clusters per cell and event in BTF data.}
\label{fig:clusters}
\end{figure}

For our 7~mm cell we expect about 12 clusters on average, given that the mean free path for cluster 
production, estimated from GARFIELD simulations, is 0.6 mm. The distribution of the number of identified clusters is shown in Figure~\ref{fig:clusters}. 
The average ($\sim$ 4) suggests an efficiency of $\sim$~33\%, to be compared with an efficiency of $\sim$~50\% obtained in the simulation.

Many effects are expected to reduce the number of identified signals, and in particular:
\begin{enumerate}
\item the gas gain fluctuations, which are typically large in a gas chamber, and can make a fraction of signals to be below the noise level -- this is expected to significantly 
improve in future applications, where better electromagnetic shielding and grounding schemes will be adopted;
\item the gain suppression induced on subsequent clusters by the space charge of the first avalanches -- a reduced noise level will also allow to recover part of this
inefficiency;
\item the shape of the single-cluster signal, namely its finite rise and fall times, originating from both the time development of the avalanche and the limited 
bandwidth of the readout electronics, which prevents to resolve very close clusters.
\end{enumerate}
%
%da dove viene il 0.6? inoltre quali e quanti dati abbiamo usato?
 
\section{Conclusions}

We have studied the single cluster response of a drift chamber prototype to 8~keV X-rays at the XLab Frascati of the INFN Frascati's National Laboratories. 
We obtained the response of the chamber and the electronics to single clusters, a fundamental input for developing refined cluster counting/timing techniques. 
We also developed a method to measure the chamber gain in our configuration. % commento sul punto di lavoro
Thanks to these data, a novel algorithm for cluster identification, based on a Wiener filter, has been developed and preliminarily tested.

\section{Acknowledgments}
We thank our colleagues of the MEG-II collaboration for the kind support in the construction and operation of the drift chamber prototype used for this study. 
S.~Dabagov would like to acknowledge the support by the Ministry of Education and Science of RF in the frames of Competitiveness Growth Program of 
NRNU MEPhI, Agreement 02.A03.21.0005. F.~Renga and C.~Voena acknowledge the support from MIUR (grant RBFR138EEU\_001/I18C13000090001).


\begin{thebibliography}{9}

\bibitem{degrad} \href{http://consult.cern.ch/writeup/magboltz/}{\tt http://consult.cern.ch/writeup/magboltz/}.
\bibitem{Wiener} Wiener H, Extrapolation, Interpolation, and Smoothing of Stationary Time Series, John Wiley and Sons, Inc., New York 1949.
\bibitem{UpgradeProposal} Baldini A M et al., MEG Upgrade Proposal, \href{http://arxiv.org/abs/1301.7225}{\tt arXiV:1301.7225 (2013)}.
\bibitem{reso_paper} Baldini A M et al., paper in preparation
\bibitem{ClusCount} Davidenko A et al., Measurements of the relativistic increase of the specific primary ionization in a streamer chamber, \emph{Nucl.~Instr.~and~Meth.} {\bf 67} (1969) 325-330.
\bibitem{ClusTime} Tassielli G et al., Improving spatial resolution and particle identification, \emph{Nucl.~Instr.~and~Meth.} {\bf A572} (2007) 198-200.
\bibitem{Xlab} \href{http://www.lnf.infn.it/xlab/index.html}{\tt http://www.lnf.infn.it/xlab/index.html}.
\bibitem{Rolandi} Blum W, Riegler W and Rolandi L, Particle Detection with Drift Chamber, Springer-Verlag, Berlin Heidelberg, 2008.
\bibitem{Mobility} Lindinger W and Albritton D L, Mobilities of various mass-identified positive ions in helium and argon, \emph{The Journal of Chemical Physics} 62 (1975) 3517.
\bibitem{TSpectrum} Morhac M et al., Identification of peaks in multidimensional coincidence gamma-ray spectra, \emph{Nucl.~Instr.~and~Meth.} {\bf A443} (2000) 108-125; 
the implementation within the ROOT data analysis framework has been used.

\end{thebibliography}
\end{document}